TITLE PAGE

# Comparison of dynamic behavior of ferroelectric and ferromagnetic nematic suspensions

**Authors**

N. Sebastian,[a] D. Lisjak,[a] M. Čopič,[a,b], O. Buchnev[c], and A. Mertelj[a,*]

**Affiliations**

[a]J. Stefan Institute, Ljubljana, Slovenia.
[b]Department of Physics, Faculty of Mathematics and Physics, University of Ljubljana, Ljubljana, Slovenia.
[c] Optoelectronics Research Centre and Centre for Photonic Metamaterials, University of Southampton, Southampton, SO17 1BJ, U.K.

*Corresponding author: Alenka Mertelj, alenka.mertelj@ijs.si

**Abstract:** The pioneering realization of suspensions of ferroelectric nanoparticles in a nematic host was one of excellent contributions of Yuri Reznikov to the science of liquid crystals [Y. Reznikov et al., Appl. Phys. Lett. 82 (2003) 191]. This achievement created great excitement as entailed the enhancement of materials dielectric properties and increase of the phase transition temperature between nematic and isotropic phase. In this contribution, we examine the spectrum of fundamental fluctuations in one of his ferroelectric suspensions by dynamic light scattering measurements, which gives insight into the coupling between the particles and the orientation of the liquid crystalline matrix. We set side by side these results with the equivalent ones obtained for the case of stable suspensions of ferromagnetic nanoplatelets in a nematic matrix showing macroscopic ferromagnetic ordering. The underlying origin of the difference between the particle-matrix coupling observed in both cases is discussed and its effect on the orientational fluctuations spectrum is compared.

Abbrevations

LC – liquid crystal

NLC – nematic liquid crystal

FM – ferromagnetic

FE – ferroelectric

DLS – dynamic light scattering

DDM – differntial dynamic microscopy

## 1   Introduction

A variety of different composites of nano and microparticles and liquid crystals (LCs) has been studied in the last two decades[1]. Among those, suspensions of ferroic nanoparticles in nematic liquid crystals (NLCs) are particularly interesting as they have many interesting properties as was recently reviewed by Reznikov et al.[2] Most remarkable is the influence of ferroic nanoparticles in the way the liquid crystal material responds to external fields. LCs are orientationally ordered fluids, in which elongated molecules on average orient along a common direction, called director. Their most important and exploited property is the possibility of controlling their orientation, and thus their optic response, using external fields. While the orientation of pure LCs can be controlled with small electric fields, the response to magnetic fields is much weaker. The idea to enhance magnetic sensitivity by doping nematic LCs (NLCs) with ferromagnetic(FM) particles is very old[3] and was followed by many experimental studies. The first stable suspension of needle-like FM particles in a NLC with large magneto-optic response was produced by Amer et al[4]. However, it was not till much later that the pioneering work by Reznikov et al.[5] led to the realization of the first suspension of ferroelectric (FE) particles in NLC. This achievement attracted significant attention since the ferroelectric suspensions showed an increase of nematic-isotropic phase transition temperature[6] and enhanced dielectric properties[5,7–9]. In the years 2005-2007 Prof. Reznikov kindly provided suspensions of FE nanoparticles to two authors (M. Č. and A. M.) of this paper to study dynamic properties of the materials[10]. In this paper we are revisiting some of unpublished experimental results obtained during this collaboration and comparing them to recent experiments in the suspensions of FM nanoplatelets in NLCs.

While most of the suspensions of FM nanoparticles exhibit paramagnetic behavior, it has been shown that the FM nanoplatelets suspended in a NLC can exhibit ferromagnetic ordering[11,12]. In the case of FE suspensions observation of possible formation of ferroelectric ordering of the particles is difficult because of ionic impurities, which screen the electric field caused by the electric dipoles of the FE nanoparticles. In principle, when ferroelectric ordering is present, the response of the material should depend on the sign of the applied DC electric field. Asymmetry in the response to positive and negative voltages was observed by Cook at al[13]. The preparation of feroic suspensions, their static properties, the advantages and the challenges were reviewed in Ref. [2].

The aim of this paper is comparison of the dynamics of the orientational fluctuations in suspensions of ferroelectric and ferromagnetic nanoparticles in nematic liquid crystals (NLC), which exhibit either ferroelectric or ferromagnetic ordering. The collective ordering of the nanoparticles, coupled to the orientation of the NLC, is reflected also in the spectrum of the eigenmodes of the orientational fluctuations of the NLC[10]. We will focus on the observation and the description of the behavior in the absence of external fields, in which case free charges/ions do not play a significant role. The paper is organized as follows. The suspensions and experimental setups are described in Materials and methods and Experimental. A macroscopic description of the suspensions and the calculation of orientational fluctuations is given in the section Theory. Finally, the results of dynamic light scattering and differential dynamic microscopy experiments in FE and FM suspensions are presented and discussed in the sections Results and Discussion.

## 2   Materials and methods

The suspension of FE particles made of $Sn_2P_2S_6$ in NLC LC18523 (Merck) was prepared as described in Ref. [10]. In brief, $Sn_2P_2S_6$ nanoparticles, prepared by milling resulting in the particles' size of 10-50 nm, were suspended in the liquid crystal mixture LC18523 ($T_{NI}$ = 55 °C). The spontaneous polarization of bulk $Sn_2P_2S_6$ is 0.14 As/m$^2$[14] and the volume concentration of the particles in the suspension was about 3·10⁻³. The suspension in the isotropic phase was filled in the homemade LC cells, which induced perpendicular (homeotropic) orientation of the NLC.

The suspension of FM platelets consisting of scandium substituted $BaSc_xFe_{12-x}O_{19}$ in NLC E7 (Merck) was prepared as described in Ref. [15]. In brief, the $BaSc_xFe_{12-x}O_{19}$ nanoplatelets were synthesized hydrothermally[16] and suspended in the liquid crystal mixture E7 ($T_{NI}$ = 58 °C). The thickness of the platelets is 5 nm and the distribution of the platelet diameter is approximately log-normal, with a mean of 70 nm and a standard deviation of 38 nm. The volume concentration of the platelets is about 1.3·10⁻³. The suspension was filled in LC cells with rubbed surfaces (thickness 20 μm, Instec Inc.), which induced homogeneous in plane orientation of NLC. During the filling the magnetic field of 8 mT was applied along the rubbing direction, so that the samples were magnetically monodomain with the magnetization **M** of 200 A/m parallel to the NLC orientation.

Polarized dynamic light scattering (DLS) and differential dynamic microscopy (DDM) were used to study the dynamics of orientational fluctuations as described in Refs. [10,15]. In both cases autocorrelation function of scattered light intensity $g_2$ is measured, from which relaxation rates of the fluctuations eigenmodes can be obtained.

## 3  Experimental

In DLS experiment we used a standard setup, using a frequency-doubled diode-pumped ND:YAG laser (532 nm, 80 mW) and ALV-6010/160 correlator to obtain the autocorrelation function of the scattered light intensity. The direction and the polarization of the incoming and detected light were chosen so that pure twist mode was observed, that is the scattering vector **q**$_s$ was perpendicular to the nematic director and, the polarizations of incoming and scattered light were ordinary and extraordinary, respectively. A single mode optical fiber with a GRIN lens was used to collect the scattered light within one coherence area. We fitted the intensity autocorrelation function $g_2$ with $g_2 = 1 + 2(1-j_d)j_d g_1 + j_d^2 g_1^2$ where $j_d$ is the ratio between the intensity of the light that is scattered inelastically and the total scattered intensity, and was either a single, $g_1 = \text{Exp}(-t/\tau)$, or double exponential function, $g_1 = a_1 \text{Exp}(-t/\tau_1) + (1-a_1)\text{Exp}(-t/\tau_2)$. The relaxation rate $1/\tau$ was attributed to the chosen eigenmode of orientational fluctuations with the wavevector **q** equal to the scattering vector **q**$_s$.

DDM is an alternative method to DLS. It has been demonstrated that it can be used in NLCs to measure relaxation rates of orientational fluctuations in the bulk[17] and in confinement[15,18]. Its advantage is that it can also probe slow dynamics at small wavevectors where usual DLS experiments are difficult to perform. We used Nikon Optihot 2 - POL microscope with SLWD objective (20x, NA=0.35). The light source of the microscope was a halogen lamp (Osram HLX64610), the numerical aperture of the condenser was set to 0.1. The polarization of the incoming light was parallel to the direction of the nematic director. The analyzer was parallel to the polarizer. A sequence of 15000 images with resolution 512x512 pixels and frame rate of 50 or 100 images per second was taken with CMOS camera (IDS Imaging UI-3370CP). It can be shown that in such a geometry the twist mode can be measured when the in-plane component of the wavevector is perpendicular to the director. The calculated intensity autocorrelation function $g_2^{DDM}$ is in this case related to $g_1$ by $g_2^{DDM}$ = A $g_1$ +B.

All measurements were performed at room temperature.

# 4 Theory

## 4.1 Order parameters of the suspensions

The orientation of NLCs is denoted by the director **n**, which is a unit vector with inversion symmetry $\mathbf{n} \equiv -\mathbf{n}$. The order parameter of the phase is a traceless tensor $\mathbf{Q} = S(\mathbf{n} \otimes \mathbf{n} - \frac{1}{3}\mathbf{I})$, where $S$ is the scalar order parameter describing how well the molecules are on average oriented along **n**[19]. At constant temperature (and away from defects) $S$ is constant, and the nematic order can be described only by **n**. If the dipole moments of the suspended nanoparticles exhibit FE or FM order (Figure 1), the order parameter which describes the orientation of the dipole moments is either polarization **P** or magnetization **M**, which are volume densities of the dipole moments.

Macroscopically, the suspensions can be therefore described by two order parameters, the nematic tensor order parameter **Q** and, in the FE case the polarization $P_0 \mathbf{n_p}$ or in the FM case the magnetization $M_0 \mathbf{m}$, where $\mathbf{n_p}$ and **m** are unit vectors which denote orientation of the electric polarization and the magnetization, respectively. $M_0$ and $P_0$ are the magnitudes of the magnetization and the polarization, respectively. If we assume uniform concentration of the particles, constant temperature, constant $S$, $P_0$, and $M_0$, the free energy density can be written

$$f = f_0 + f_{elast} + f_{field} + f_{coupl} \tag{1}$$

where $f_0$ is the free energy describing the ferroelectric/ferromagnetic nematic phase. It includes only $P_0$, or $M_0$ and the scalar order parameter of the nematic phase $S$, which are at constant temperature assumed to be constant. The term $f_{elast}$ is the usual Frank elastic energy

$$f_{elast} = \tfrac{1}{2} K_1 (\nabla \cdot \mathbf{n})^2 + \tfrac{1}{2} K_2 (\mathbf{n} \cdot (\nabla \times \mathbf{n}))^2 + \tfrac{1}{2} K_3 (\mathbf{n} \times (\nabla \times \mathbf{n}))^2 , \tag{2}$$

where $K_i$ ($i$ = 1,2,3) are splay, twist and bend elastic constant, respectively. $f_{field}$ describes the coupling of a polar order with external fields, $-P_0 \mathbf{n}_p \cdot \mathbf{E}$ or $-\mu_0 M_0 \mathbf{m} \cdot \mathbf{H}$, and the coupling of the nematic order with external fields, $-\varepsilon_a \varepsilon_0 (\mathbf{n} \cdot \mathbf{E})^2$ or $-\chi_a \mu_0 (\mathbf{n} \cdot \mathbf{H})^2$, where $\mu_0$ is the vacuum permeability and $\varepsilon_0$ the vacuum permittivity. The dielectric anisotropy $\varepsilon_a$ and the magnetic anisotropy $\chi_a$ are the differences of dielectric constant and magnetic susceptibility along and perpendicular to **n**, respectively.

In the following subsection we will focus on the coupling term $f_{coupl}$, which describes interaction between the orientation of **n** and $\mathbf{n_p}$ (or **m**).

## 4.2 Coupling between nematic orientation and the nanoparticles

There is a crucial difference on how ferromagnetic and how ferroelectric particles couple to orientation of the liquid crystalline matrix. While the magnetic field which surrounds the magnetic nanoparticles has a little effect on the LC orientation as the magnetic anisotropy is small ($\chi_a \approx 10^{-6}$), the electric field around the ferroelectric particles is strongly coupled to LC (the dielectric anisotropy of NLCs typically takes values about 10). The magnetic moments of the particles are of the order of $10^{-20}$-$10^{-18}$ Am$^2$, which produces the field of less than 1 mT at a distance of 50 nm from the particle center. On the other hand ferroelectric particles have electric dipole moments of the order of $10^{-25}$ - $10^{-23}$ As.m, which gives the electric field of $10^5$ - $10^7$ V/m at the same distance. Therefore, in the magnetic case, the term in the free energy density corresponding to the coupling of the field due to the dipole moment of the particles with the nematic director, i.e., $-\tfrac{1}{2} \chi_a \mu_0 (\mathbf{n} \cdot \mathbf{H}_{dip})^2$, is more than ten orders of magnitude smaller than in the electric case, i.e., $-\tfrac{1}{2} \varepsilon_a \varepsilon_0 (\mathbf{n} \cdot \mathbf{E}_{dip})^2$.

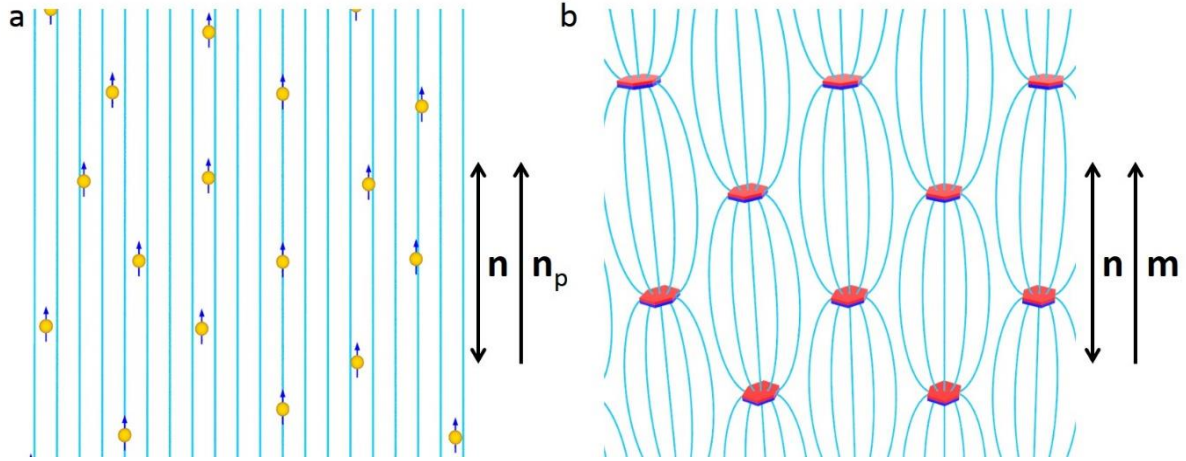

*Figure 1: Schematic presentation of the ordered ferroic nanoparticles embedded in LC matrix. A) Small spherical particles (yellow) do not significantly disturb the orientation of LC (light blue). In the case of ferroelectric particles, the origin of the coupling between the orientation of the dipoles (blue arrows) and LC director is in the interaction between the LC and the electric field caused by the electric dipoles of the particles. B) Larger anisotropic particles distort the orientation of the LC around them to a greater extend. In the case of magnetic nanoplatelets made of BaHF with homeotropic anchoring at their surface, the platelets orient with their short axis, i.e., magnetic dipole moments, along the director. The origin of the coupling in this case is the anchoring of the LC molecules at the surface of the platelets. Magnetic moments of the platelets are denoted by red (N) and blue (S) colors. Arrows **n**, **n**$_p$, and **m** denote the orientation of the NLC director, the electric polarization, and the magnetization, respectively.*

Nanoparticles are coupled to liquid crystal orientation also through the anchoring of the liquid crystalline molecules on the surface of the particles. The surface anchoring and the nanoparticle's shape determine how an anisotropic nanoparticle is oriented with the respect to **n**. If the LC is reoriented, the nanoparticle orientation will also follow and, vice versa, if the particle is reoriented the orientation of the surrounding LC will also follow. In the latter case, a collective reorientation of the LC can be achieved if the nanoparticles' concentration is high enough[3,20]. In theory this works only for anisotropic particles, however, it was observed that also spherical nanoparticles can cause reorientation of the LC matrix and that was attributed to memory effect of the anisotropic anchoring [21–23]. The effect is however much smaller. Indeed, an effective anchoring strength of NLC with magnetite nanorods was one order or magnitude higher then with spherical nanoparticles[24]. In the case of suspensions of FM particles in the NLC, this surface anchoring constitutes an indirect way of coupling between the magnetic field and the NLC director. It works only if the direction of particles' magnetic moments are pinned to the particle shape, i.e., if the magnetic field causes mechanical reorientation of the particles (Brownian relaxation) and not only reorientation of the magnetic moments within the particles (so called Neel relaxation). On the contrary, in the case of FE particles, the coupling will exist whether the particles mechanically reorient with the reorientation of the electric dipole moments or not.

The additional terms in the free energy of the system due to the coupling between the orientation of **n** and orientation of the particles is in both cases similar and can be written

$$f_{coupl,E} = \tfrac{1}{2} g_{el} \left( \mathbf{n} \cdot \mathbf{n}_p \right)^2 \tag{3}$$

$$f_{coupl,M} = \tfrac{1}{2} g_m \left( \mathbf{n} \cdot \mathbf{m} \right)^2 \tag{4}$$

The coupling parameters $g_{el}$ and $g_m$ depend on the concentration of the particles, the electric/magnetic dipole moments, the size, the shape, and the surface properties of the particles. As described above $g_{el}$ is determined by the coupling of **n** with the electric field produced of the electric dipole of the particles, whereas the main contribution to $g_m$ comes from the surface anchoring of the LC molecules at the surface of the particles. Thus, the coupling is much stronger in the electric case with $g_{el}$ of the order of $10^3 – 10^4$ J/m$^3$, whereas in the magnetic case the largest effect was found for the platelets, in which case it is about 5 J/m$^3$.

### 4.3 The dynamics of orientational fluctuations

In the ordinary nematic liquid crystals orientational fluctuations are fundamental thermal excitations of the director field. In the bulk NLC with uniform **n** their eigenmodes are plane waves with relaxation rates that depends on the elastic constants and viscosities of the material and the wave vector **q**, and can be written as $\delta \mathbf{n} = \delta \mathbf{n}_0 \mathrm{Exp}(-i\mathbf{q}\cdot\mathbf{r}) + \mathrm{c.c.}$. **n** is a unit vector, so it has two degrees of freedom, which results in two branches of the fluctuations eigenmodes, splay-bend and twist-bend branch with the relaxation waves given by[19]

$$\frac{1}{\tau_\beta(\mathbf{q})} = \frac{K_\beta q_\perp^2 + K_3 q_\parallel^2}{\eta_\beta(\mathbf{q})} \tag{5}$$

where $\beta = 1$ denotes splay-bend mode and $\beta = 2$ twist-bend mode, and $q_\perp$ and $q_\parallel$ denote the components of wave vector **q** perpendicular and parallel to **n**, respectively. The effective viscosities are

$$\eta_1(\mathbf{q}) = \gamma_1 - \frac{\left(q_\perp^2 \alpha_3 - q_\parallel^2 \alpha_2\right)^2}{q_\perp^4 \eta_b + q_\perp^2 q_\parallel^2 (\alpha_1 + \alpha_3 + \alpha_4 + \alpha_5) + q_\parallel^4 \eta_c}$$

$$\eta_2(\mathbf{q}) = \gamma_1 - \frac{\alpha_2^2 q_\parallel^2}{q_\perp^2 \eta_a + q_\parallel^2 \eta_c} \tag{6}$$

where $\alpha_i$ are Leslie viscosity coefficients, $\gamma_1$ is rotational viscosity and $\eta_{a,b,c}$ Miesowicz viscosities[19]. In general, $\eta_{1,2}$ are smaller than $\gamma_1$ because of the coupling between **n** and the flow, so called backflow effect[19]. However, in the case of pure twist mode ($q_\parallel = 0$), there is no backflow and $\eta_2 = \gamma_1$.

In the suspensions, an additional order parameter, coupled to **n**, is present, which leads to the enrichment of the spectrum of the orientational fluctuations[10,11]. The fluctuations eigenmodes are again plane waves and can be written as $\{\delta\mathbf{n}, \delta\mathbf{n}_p\} = \{\delta\mathbf{n}_0, \delta\mathbf{n}_{p0}\}\mathrm{Exp}(-i\mathbf{q}\cdot\mathbf{r}) + \mathrm{c.c.}$ ( or $\{\delta\mathbf{n}, \delta\mathbf{m}\} = \{\delta\mathbf{n}_0, \delta\mathbf{m}_0\}\mathrm{Exp}(-i\mathbf{q}\cdot\mathbf{r}) + \mathrm{c.c.}$ ). The number of the degrees of freedom in this case is four, which results in the splitting of each eigenmode of the fluctuations observed in pure NLC into two eigenmodes. The relaxation rates can be calculated from linearized dynamic equations for **n**, **n_p** (or **m**) and, the flow velocity **v**. Only the most simple case, that is pure twist mode in which **v** = 0, will be considered here. In this case, the fluctuations are perpendicular to the wave vector $(\delta\mathbf{n}, \delta\mathbf{n}_p$ (or $\delta\mathbf{m})) \perp \mathbf{q}$ and to the initial orientation of **n** (Figure 2). The dynamics is governed by the balance equations, in which the molecular fields $\mathbf{h}_n$ and $\mathbf{h}_{np}$ (or $\mathbf{h}_m$) drive the quasi currents $\dot{\mathbf{n}}$ and $\dot{\mathbf{n}}_p$ (or $\dot{\mathbf{m}}$). In the case of FE suspension and in the absence of external fields, two linearized scalar equations for the twist fluctuations are obtained:

$$\delta\dot{n}_2 = -\frac{1}{\gamma_1}\left(K_2 q_\perp^2 \delta n_2 + (\delta n_2 - \delta n_{p2})g_{el}\right) \tag{7}$$

$$\delta\dot{n}_{p2} = -\frac{1}{\gamma_{FE}}(\delta n_{p2} - \delta n_2)g_{el} \qquad (8)$$

which have two solutions of the type $\{\delta n_2, \delta n_{p2}\} = \{\delta n_{20}, \delta n_{p20}\}\text{Exp}(-t/\tau)\text{Exp}(-i\mathbf{q}\cdot\mathbf{r})$ with the relaxation rates

$$\frac{1}{\tau_{\pm}} = \frac{K_2 q_{\perp}^2}{2\gamma_1} + \frac{g_{el}(\gamma_1 + \gamma_{FE})\left(1 \pm \sqrt{1 - \frac{2K_2 q_{\perp}^2 \gamma_{FE}(\gamma_1 - \gamma_{FE})}{g_{el}(\gamma_1 + \gamma_{FE})^2} + \frac{K_2^2 q_{\perp}^4 \gamma_{FE}^2}{g_{el}^2(\gamma_1 + \gamma_{FE})^2}}\right)}{2\gamma_1 \gamma_{FE}} \qquad (9)$$

If the condition $K_2 q_{\perp}^2 \ll g_{el}$ is considered, they simplify to

$$\frac{1}{\tau_+} \approx \frac{g_{el}(\gamma_1 + \gamma_{FE})}{\gamma_1 \gamma_{FE}} + \frac{K_2 q_{\perp}^2 \gamma_{FE}}{\gamma_1(\gamma_1 + \gamma_{FE})} \qquad (10)$$

$$\frac{1}{\tau_-} \approx \frac{K_2 q_{\perp}^2}{\gamma_1 + \gamma_{FE}} \qquad (11)$$

Here $\gamma_{FE}$ is the rotational viscosity for $\mathbf{n_p}$, analogous to $\gamma_1$. At $q_{\perp} = 0$ the relaxation rate of the first mode is finite, which is typical for an optic mode, whereas the relaxation rate of the second mode quadratically depends on $q_{\perp}$, which is a characteristics of a hydrodynamic mode. As represented in Figure 2a, in the optic mode $\mathbf{n}$ and $\mathbf{n_p}$ fluctuate in counter phase, whereas in hydrodynamic case they fluctuate in phase (Figure 2a).

On the other hand, it has been recently shown that in the suspensions of FM platelets in NLCs also a dynamic dissipative coupling between $\mathbf{n}$ and $\mathbf{m}$ exists[25]. Taking this dynamic coupling into account, the dynamic equations for the twist mode in the FM case are

$$\delta\dot{n}_2 = -\frac{1}{\gamma_1}\left(K_2 q_{\perp}^2 \delta n_2 + (\delta n_2 - \delta m_2)g_m\right) - \chi_2(\delta m_2 - \delta n_2)g_m \qquad (12)$$

$$\delta\dot{m}_2 = -\frac{1}{\gamma_{FM}}(\delta m_2 - \delta n_2)g_m - \chi_2\left(K_2 q_{\perp}^2 \delta n_2 + (\delta n_2 - \delta m_2)g_m\right), \qquad (13)$$

where $\chi_2$ is dynamic dissipative coupling coefficient. The relaxation rates then can be calculated

$$\frac{1}{\tau_{\pm}} = \frac{K_2 q_{\perp}^2}{2\gamma_1} + \frac{g_m(\gamma_1 + \gamma_{FM} - 2\chi_2\gamma_1\gamma_{FM})\left(1 \pm \sqrt{1 + \frac{2K_2 q_{\perp}^2 \gamma_{FM}(\gamma_1 - \gamma_{FM} - 2\chi_2\gamma_1\gamma_{FM} + 2\gamma_1^2\gamma_{FM}\chi_2^2)}{g_m(\gamma_1 + \gamma_{FM} - 2\chi_2\gamma_1\gamma_{FM})^2} + \frac{K_2^2 q_{\perp}^4 \gamma_{FM}^2}{g_m^2(\gamma_1 + \gamma_{FM} - 2\chi_2\gamma_1\gamma_{FM})^2}}\right)}{2\gamma_1 \gamma_{FM}}. \qquad (14)$$

Again, if the condition $K_2 q_{\perp}^2 \ll g_m$ is taken into account, Equation (14) can be simplified to

$$\frac{1}{\tau_+} \approx \frac{g_m(\gamma_1 + \gamma_{FM} - 2\chi_2\gamma_1\gamma_{FM})}{\gamma_1 \gamma_{FM}} + \frac{K_2 q_{\perp}^2 \gamma_{FM}(\gamma_1\chi_2 - 1)^2}{\gamma_1(\gamma_1 + \gamma_{FM} - 2\chi_2\gamma_1\gamma_{FM})} \qquad (15)$$

$$\frac{1}{\tau_-} \approx \frac{K_2 q_{\perp}^2 (1 - \gamma_1\gamma_{FM}\chi_2^2)}{\gamma_1 + \gamma_{FM} - 2\chi_2\gamma_1\gamma_{FM}} \qquad (16)$$

Analogously to the FE case, the first mode corresponds to an optic mode and the second one to a hydrodynamic mode.

## 5 Results

### *5.1 Orientational fluctuations in FE case*

Using DLS we studied the spectrum of relaxations of the twist fluctuations and its dependency on $q_\perp$ in the FE suspensions. Two modes were observed. The relaxation rate of the slower mode was about a factor of 0.7 smaller than that observed in the pure NLC. The faster mode, which was only observed in the suspensions but not in the pure NLC, did not depend on the wave vector and its relaxation rate was about 2 orders of magnitude larger than that of the slower mode. On the contrary, its amplitude was much smaller than that of the slower mode. The faster mode was observed in the scattering geometry in which polarizer and analyzer were perpendicular to each other and to **n**. In this geometry, the director fluctuations in the pure LC were not observed, whereas in the suspensions they were observed at small scattering vectors, but the scattering amplitude was very small. We fittted the relaxation rates using expressions (10) and (11). The fits (lines in Figure 2b) yield $g_{el} = \gamma_1 \cdot 55.6$ kHz $\approx 8900$ J/m$^3$ and $\gamma_{FE} = 0.45\gamma_1$ for $\gamma_1 = 0.16$ Pa.s, which is an extrapolated value for $T = 27°C$ obtained from $\gamma_1$ (20°C) = 0.29 Pa.s (Merck value) and temperature dependence of the ratio $K_2/\gamma_1$ in the pure LC 18532[10].

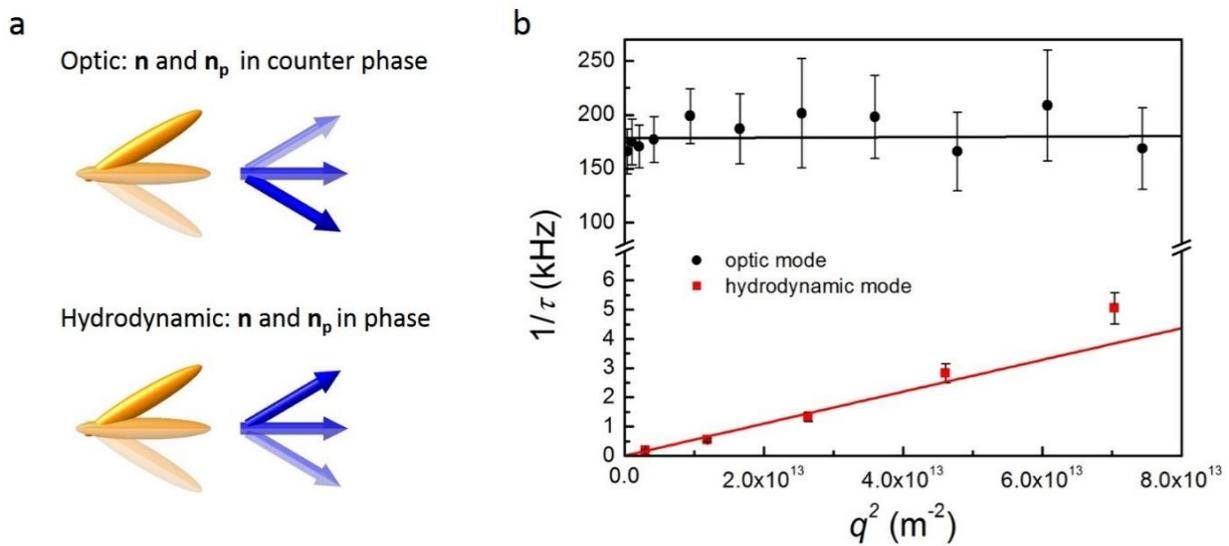

Figure 2: Twist fluctuation modes observed in suspension of FE nanoparticles ($Sn_2P_2S_6$) in nematic liquid crystal LC 18532. a) Scheme of fluctuations of **n** (yellow ellipsoids) and **n**$_p$ (blue arrows) in optic and hydrodynamic modes. In the first, **n** and **n**$_p$ fluctuate in counter phase, whereas in the second they fluctuate in phase. B) Measured dispersion curves for optic (black circles) and hydrodynamic (red squares) modes. Lines are the fits. For the details, see text.

## 5.2 Orientational fluctuations in FM case

The dependence of the twist relaxation rate on $q_\perp$ in the FM suspensions was measured by DDM experiments. In this case we observed only one mode with a relaxation rate very close to the one observed in the bulk (Figure 3). Comparing with the FE case, the coupling coefficient $g_m$ is much smaller, and thus the splitting between the modes is small, making it difficult to resolve them separately. We calculated the relaxation rates using expression (14) and values for $g_m = 4$ J/m$^3$, $\chi_2 = 4$ (Pa.s)$^{-1}$, $\gamma_1 = 0.14$ Pa.s and, $\gamma_{FM} = 0.27$ Pa.s, which we obtained from the static and dynamic responses on magnetic field in a similar way as described in References [15,25]. The calculated splitting between the modes is indeed small (lines in Figure 3) and the measured values lie between them.

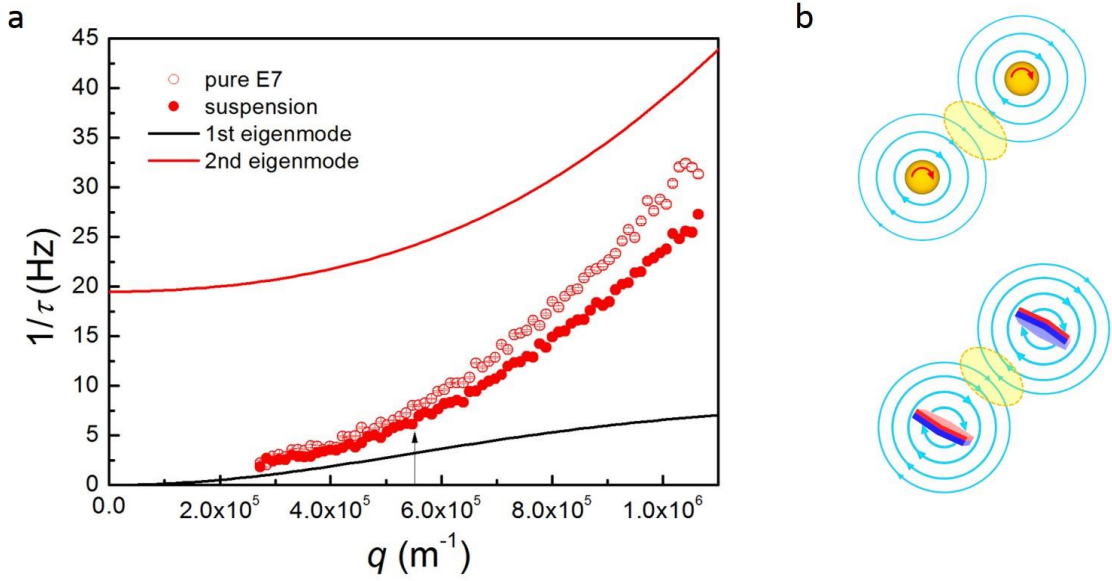

Figure 3: a) Comparison of the twist fluctuation modes observed in pure nematic liquid crystal E7 and in suspension of FM nanoplatelets (BaSc$_{0.5}$Fe$_{11.5}$O$_{19}$) in E7. Lines are calculated relaxation rates using expression (14). For the details, see text. b) Schematic of flow (blue lines) around rotating spheres and platelets. In the regions denoted yellow additional dissipation occurs.

## 6 Discussion

The results show that information about the coupling of the order parameters in the FE and FM suspensions can be obtained from the analysis of the modes of orientational fluctuations. In the FE case, the coupling is strong which results in two modes with the relaxation rates with values values that differ by two orders of magnitude and can be clearly separated from each other in the scattering experiment. From the measurements we were able to determine the coupling constant $g_{el}$ and the FE rotational viscosity, which govern the collective orientational dynamics of the FE particles. The coupling constant can be estimated by assuming that the electric field caused by the ferroelectric particles

$$E_p \sim \frac{\phi P}{\varepsilon \varepsilon_0} \quad (17)$$

is coupled to **n**

$$f_{coupl} = \tfrac{1}{2} \varepsilon_a \varepsilon_0 \left( \mathbf{n} \cdot \mathbf{E}_p \right)^2 = \frac{\varepsilon_a P^2 \phi^2}{2\varepsilon^2 \varepsilon_0} \left( \mathbf{n} \cdot \mathbf{n}_p \right)^2 \quad (18)$$

This gives $g_{el}$ of 1000 J/m³ (for LC 18523 $\varepsilon_a$ = 2.7, $\varepsilon$ = 7), which is significantly less than what we measured. Similarly, we can estimate the FE rotational viscosity by assuming that the dissipation comes from independent contribution of individual spheres with rotational friction coefficient, $8\pi\eta R^3$, and the number density $N/V$: $\gamma_{FM} \approx 8\pi\eta R^3 N/V = 6\eta\phi$, so $\gamma_{FE}/\gamma_1 = 0.018\eta/\gamma_1$, where $\eta$ is shear viscosity of the NLC and is comparable to $\gamma_1$. This value is an order of magnitude smaller than the one obtained from the measurements. The reason for this discrepancy can lay in our assumption that in the suspension the ratio $K_2/\gamma_1$ is the same as in the pure NLC, which would be true if the first would consist only of NLC and the particles. However, in the suspension, there are also surfactant molecules and maybe some residual solvent present, which could affect the ratio $K_2/\gamma_1$. Here, it should be noted that this ratio does not depend on the order parameter $S$, because both, $K_2$ and $\gamma_1$ are proportional to $S^2$. If we assume that the main contribution to the reduction of this ratio is due to impurities, that is, $\gamma_{FE}/\gamma_1$ is small, e.g. $\gamma_{FE}/\gamma_1 = 0.02$, then $g_{el} = 560$ J/m³, which is much closer to the estimated value. It is also possible that in the FE case dynamic coupling between the order parameters additionally exists transforming the relaxation rates into expressions analogous to Expressions (15) and (16). For $g_{el} = 1000$ J/m³, that yields $\gamma_{FE}/\gamma_1 \approx 0.03$ and $\chi_2 \approx 0.6$ (Pa.s)$^{-1}$.

The origin of the light scattering by orientational fluctuations in NLCs is the anisotropy of index of refraction. Optically NLCs are uniaxial materials with the optical axis parallel to **n**, so the fluctuations $\delta$**n** are fluctuations of the optical axis. In the suspensions, light is also scattered by the nanoparticles. Similarly as in the suspension in isotropic solvents, the rotational motion can be observed in the depolarized geometry, that is when analyzer and polarizer are perpendicular to each other, if the particles are optically anisotropic. A particle in NLC causes optically anisotropic deformation, which scatters light, so collective fluctuations of the orientation of particles, i.e., $\delta$**n**$_p$ or $\delta$**m**, also contributes to the scattered intensity. This is probably the reason why we observe the optic mode in the scattering geometry in which the director fluctuations in pure NLCs are not observed.

Although the situation in FM suspensions is theoretically analogous to FE case, the experimentally determined parameters are quite different. The advantage of the magnetic case is that the parameters can be measured in other experiments, and so we can predict the relaxation rates of the fluctuation modes. The disadvantage is that due to the small coupling, the relaxation rates of the optic and the hydrodynamic modes are close to each other, so that we can not distinguish them. The obtained relaxation rate is thus a weighted average of the two modes. The weights depends on the amplitudes of the modes, i.e., $\{\delta n_{20}, \delta m_{20}\}$ and on their scattering cross sections. The latter are known for $\delta n_2$[19], but not for $\delta m_2$. As we discussed in the previous paragraph, the origin of the scattering on $\delta$**m** is the scattering on the particles and on the deformation of the NLC around them. We can estimate FM orientational viscosity in a similar way as we did for FE viscosity. Using the friction coefficients for disks[26], we get for independent platelets $\gamma_{FM}/\gamma_1 \approx 0.032\eta/\gamma_1$. This is much smaller than the measured value, $\gamma_{FM}^{meas.}/\gamma_1 \approx 1.9$. The assumption of independent platelets is however very crude. Rotation of the particles in the viscous media is accompanied by flow and when the particles are close to each other the flow from neighboring particles is opposite (Figure 3b), which results in an increase of the collective viscosity. Since the flow which accompanies the rotation of the platelets is much larger than that around the spheres, the discrepancies between the measured and the estimated values for independent particles is expected to be larger in the case of the platelets.

In the FM suspension, it is possible to measure the static and dynamic coupling coefficients and viscosities from the response to magnetic field, whereas the situation in the FE case is much more complicated. There, the response to DC electric field is strongly affected by ionic impurities. It has been shown that for the description of the response of the FE suspensions to DC electric field, ions must be taken into account, which screen the electric field of the FE particles and, on the other hand, the FE particles act as ion traps and cause faster relaxation [27–29]. Thus contrary to the FM case, it is not possible to extract static or dynamic coupling parameter from the response to DC electric field measurements. Additionally, the aggregation of nanoparticles is a larger problem in the FE case, because the dipolar interaction between the spherical FE particles is much larger than between magnetic platelets, in which repulsive nematic-mediated elastic interaction further helps against aggregation[11].

## 7  Conclusions

We have shown that in the suspension of ferroic particles in a NLC, which beside nematic exhibit also FE or FM ordering, the spectrum of orientational fluctuation changes. The system can be macroscopically described by two orientational order parameters, so the number of orientational fluctuation modes doubles. We focused on the twist modes and showed that beside the hydrodynamic mode, which has a typical $q^2$ dependence at small $q$, also an optic-like mode exists, which has a finite relaxation rate at $q$ = 0. Experimentally, we were able to observe both modes in the FE suspension, whereas in the FM suspension the relaxation rates of the modes are too close to each other, so that only their average could be measured. The latter was close to the twist relaxation rate in the pure NLC.

**Acknowledgements:** NS thanks the "EU Horizon 2020 Framework Programme for Research and Innovation" for its support through the Marie Curie Individual fellowship No. 701558 (MagNem). The authors acknowledge also the financial support from and the Slovenian Research Agency (PMR, MČ, and AM research core funding No. P1-0192, DL research core funding No. P2-0089 and, AM and DL the project No. J7-8267).

The authors express their deepest acknowledgement to the late Prof. Yuri Reznikov. His knowledge and contribution to the LC science in general, and to the area of ferromagnetic and ferroelectric LC colloids in particular, is well-known in the international LC community. Yuri was the PhD supervisor of Oleksandr Buchnev, for whom Yuri was not only an advisor on scientific questions, but a confidant able to suggest solutions to a host of life-related problems. Oleksandr always remembers with fondness Yuri's recommendations. His memory will be with us for many years to come.

Conflicts of interest: none